\documentclass[10pt]{article}

\newif\ifbmaversion
\bmaversionfalse
\usepackage[utf8]{inputenc}
\usepackage[T1]{fontenc}

\ifbmaversion
    \usepackage{booktabs}
    \usepackage{calc}
    \usepackage{amssymb}  
    \usepackage{graphicx}
    \usepackage{color}
    \usepackage{verbatim}
    \usepackage{enumitem}
    \usepackage{bm}
    \usepackage{subcaption}
    \usepackage[super]{nth}
    \usepackage{multirow}

    \usepackage{amsmath}
    \usepackage[plain,noend]{algorithm2e}
    \usepackage{newtxtext}
    \usepackage[subscriptcorrection]{newtxmath}
    \usepackage{hyperref}
     \usepackage{xr}

\else
    \usepackage{calc}
    \usepackage{color}
    \usepackage{verbatim}
    \usepackage{bm}
    \usepackage[super]{nth}
    
    \usepackage{natbib}
    \usepackage{hyperref}       
    \usepackage{url}            
    \usepackage{booktabs}       
    \usepackage{amsfonts}       
    \usepackage{nicefrac}       
    \usepackage{microtype}      
    \usepackage{graphicx}  
    \usepackage{float}
    \usepackage{amsthm}
    \usepackage{mathtools}
    \usepackage{amsfonts}
    \usepackage{amssymb}
    \floatstyle{ruled}
    \usepackage{xcolor}
    \usepackage{bbm}
    \usepackage{xr}
    
    \usepackage{adjustbox}
    \usepackage{wrapfig}
    \usepackage{caption}
    \usepackage{subcaption}
    \usepackage{cprotect}
    \usepackage{etoc}
    \usepackage{titlesec}
    \usepackage{enumitem}
    
    \usepackage{tikz}
    \usetikzlibrary{arrows,automata, calc, positioning}
    \usepackage{adjustbox}
    \usepackage{wrapfig}
    
    \usepackage{authblk}
    
    \usepackage{multicol}
    \usepackage{multirow}
    \usepackage{booktabs}
    
    \usepackage{color-edits}

      \usepackage[super]{nth}
    \usepackage{multirow}

    \usepackage{amsmath}
    \usepackage[plain,noend]{algorithm2e}
    \usepackage{newtxtext}
    \usepackage[subscriptcorrection]{newtxmath}
    \usepackage{hyperref}
     \usepackage{xr}

    \addauthor{vs}{red}
\fi
\ifbmaversion
    \makeatletter
    \renewcommand{\algocf@captiontext}[2]{#1\algocf@typo. \AlCapFnt{}#2} 
    \def\@algocf@capt@plain{top}
    \renewcommand{\algocf@makecaption}[2]{%
      \addtolength{\hsize}{\algomargin}%
      \sbox\@tempboxa{\algocf@captiontext{#1}{#2}}%
      \ifdim\wd\@tempboxa >\hsize
        \hskip .5\algomargin%
        \parbox[t]{\hsize}{\algocf@captiontext{#1}{#2}}
      \else%
        \global\@minipagefalse%
        \hbox to\hsize{\box\@tempboxa}
      \fi%
      \addtolength{\hsize}{-\algomargin}%
    }
    \makeatother
    
    \newcommand{\fontarxivtable}{}
    
\else
    \theoremstyle{definition}
    \newtheorem{theorem}{Theorem}
    \newtheorem{assumption}{Assumption}
    \newtheorem{example}{Example}

    \newtheorem{lemma}{Lemma}

    \def\spacingset#1{\renewcommand{\baselinestretch}%
    {#1}\small\normalsize} \spacingset{1}

    \newcommand{\kibitz}[2]{\ifnum\Comments=1{\color{#1}{#2}}\fi}

    \renewcommand{\Pr}{\ensuremath{\mathrm{Pr}}}

    \def\balign#1\ealign{\begin{align}#1\end{align}}
    \def\balignat#1\ealign{\begin{alignat}#1\end{alignat}}
    \def\bitemize#1\eitemize{\begin{itemize}#1\end{itemize}}
    \def\benumerate#1\eenumerate{\begin{enumerate}#1\end{enumerate}}
    \newenvironment{talign}
     {\csname align\endcsname}
     {\endalign}
    \def\balignt#1\ealignt{\begin{talign}#1\end{talign}}%
    \newcommand{\tbl}[1]{\caption{\small#1}}
    \newcommand{\fontarxivtable}{\small}
    
    \newenvironment{tabnote}{\vskip7pt\par\fontsize{9}{11}\ignorespaces}{\par}
   
\fi

\def\T{{ \mathrm{\scriptscriptstyle T} }}

\let\originalleft\left
\let\originalright\right
\renewcommand{\left}{\mathopen{}\mathclose\bgroup\originalleft}
\renewcommand{\right}{\aftergroup\egroup\originalright}
\DeclareRobustCommand{\VAN}[2]{#1} 

\date{Original draft: July 2023. This draft: December 2024.}

\begin{document}

\title{Automatic Debiased Machine Learning for Covariate Shifts}

\author[1]{Victor Chernozhukov}
\author[2]{Michael Newey}
\author[1]{Whitney K. Newey}
\author[3]{Rahul Singh}
\author[4]{Vasilis Syrgkanis}
\affil[1]{\footnotesize Department of Economics, Massachusetts Institute of Technology, Cambridge, MA 02142, USA}
\affil[2]{Lincoln Laboratory, Massachusetts Institute of Technology, Lexington, MA 02421, USA}
\affil[3]{Society of Fellows and Department of Economics, Harvard University, Cambridge, MA 02138, USA}
\affil[4]{Department of Management Science and Engineering, Stanford University, Stanford, CA 94305, USA}


\maketitle
\bigskip
\begin{abstract}

We present machine learning estimators for causal and predictive parameters under covariate shift, where covariate distributions differ between training and target populations. One such parameter is the average effect of a policy that alters the covariate distribution, such as a treatment modifying surrogate covariates used to predict long-term outcomes. Another example is the average treatment effect for a population with a shifted covariate distribution, like the effect of a policy on the treated group.

We propose a debiased machine learning method to estimate a broad class of these parameters in a statistically reliable and automatic manner. Our method eliminates regularization biases arising from the use of machine learning tools in high-dimensional settings, relying solely on the parameter's defining formula. It employs data fusion by combining samples from target and training data to eliminate biases. We prove that our estimator is consistent and asymptotically normal. Computational experiments and an empirical study on the impact of minimum wage increases on teen employment—using the difference-in-differences framework with unconfoundedness—demonstrate the effectiveness of our method.


\end{abstract}

\ifbmaversion
\else
    \noindent%
\fi
\vfill
\newpage



\section{Introduction}\label{sec:intro}

Machine learning models trained on one dataset are often used to estimate parameters in another dataset with a different—or "shifted"—covariate distribution. Examples include situations where training data represent a subpopulation of a larger group or when training and estimation occur at different times, leading to varying covariate distributions. For instance, a neural network trained on one dataset may be applied to new data with different covariate distributions to estimate average outcomes.

In causal inference, regression models trained on outcome, treatment, and covariate data can estimate counterfactual averages in another dataset with a different covariate distribution. Covariate shifts are important in various settings, including the estimation of counterfactual averages and causal effects~\citep{hotz2005predicting, pearl2011transportability, belloni2017program,singh2020kernel}, as well as in prediction and classification tasks where training data differ from target data~\citep{Bahng2022arxiv, WILDS}.

We focus on estimating parameters in target data that depend on regressions from training data, addressing the bias introduced by regularization or model selection during training. We propose automatic debiased machine learning estimators for these parameters, requiring only the parameter of interest without needing a full theoretical bias correction.

Our debiased estimators are constructed by plugging a regression from the training data into the target parameter formula and adding a debiasing term. This term is the average product of a balancing weight (Riesz representer) and regression residuals in the training data. The debiasing function is estimated by linking the target data with features of the training data, using only the parameter's defining formula. While similar to estimators in \citet{chernozhukov2022automatic}, our approach differs by addressing the data fusion problem where training and target data come from different sources, are statistically independent, and have different covariate distributions.

We describe the estimator, present the underlying theory, conduct simulation experiments, and provide an empirical example evaluating the effect of minimum wage on youth unemployment.

We contribute to the literature on semiparametric models under covariate shift~\citep{robins2007comment,chen2008semiparametric, hernan2011compound, stuart2011use, pearl2011transportability, rudolph2017robust, dahabreh2023efficient, li2023efficient} by proposing a debiased, cross-fitted estimator based on approximate balance for a broad class of parameters in high-dimensional settings. Previous studies have used estimators with explicit debiasing formulas tailored for specific parameters, employing propensity scores or density ratios as balancing weights. Our approach considers a wider range of parameters and constructs balancing weights through optimization problems that promote regular solutions, thereby avoiding numerical instability common in explicit debiasing formulas.

For off-policy evaluation parameters, \citet{kallus2021optimal} propose an exact balancing method. In contrast, we study a broader class of parameters and use an approximate balancing approach. While some works calibrate regression and density ratios within specific binary response models~\citep{zhou2024doubly}, we accommodate generic machine learning methods for regression estimation and allow for general response models.

We characterize balance in high-dimensional covariate shift problems by analyzing Riesz representers. Recognizing that balancing weights (Riesz representers) are identified from two sufficient statistics, we estimate these representers without relying on specific modelling or learning strategies. Our framework addresses covariate shift by estimating one sufficient statistic from the training population and another from the target population. We then combine the Riesz estimator with a generic machine learning regression estimator using debiased estimating equations
that have a double-robustness property \citep{robins1995analysis, robins1995semiparametric}. Extending cross-fitting~\citep{zheng2011cross,belloni2010lasso} to the data-fused case achieves approximate unbiasedness and asymptotic normality under weak conditions.

Previous works on high-dimensional Riesz representers and balancing weights for policy effect estimation often assume known mappings or distributions for covariates~\citep{chernozhukov2018global, hirshberg2019augmented, chernozhukov2022automatic}.
Our method removes these stringent requirements, accommodating unknown target covariate distributions.

\section{The method and theory}\label{sec:method}

\subsection{Parameters of interest and debiased estimating equations}
The parameters of interest we consider depend on the regression 
\begin{equation*}
    \gamma_{0}(X)=E(Y|X),
\end{equation*}
where $Y$ is an
outcome variable of interest, and $X$ is a vector of regressors. The parameters of interest also depend on a vector of random variables $Z$, which will often be a shifted vector of regressors with the same dimensions as $X$ but from a different distribution. Our setup applies to settings where $\gamma_{0}(X)$ is a
nonparametric regression for training data and the parameter depends on target
data $Z$ in the following way:
\begin{equation*}\theta_{0}=E\{m(Z,\gamma_{0})\},
\end{equation*}
where $m(Z,\gamma)$ is a score that depends on $Z$ and linearly depends on a possible regression $\gamma$. 
A leading example has $m(Z,\gamma)=\gamma(Z)$, giving
$$
\theta_{0}=E\{\gamma_{0}(Z)\},
$$
so that  $\theta_{0}$ is the expectation of the outcome, when the regressor distribution is shifted from that of $X$ to that of $Z$, and the regression function remains the same. 
For example, if $Y$ is a binary variable for classification into one of two groups, then $\theta_{0}$ is the classification probability for the target population.
\begin{example}[Policy effect from covariate shifting]Consider evaluating policies or treatments that affect outcomes indirectly by changing covariates. A key example is assessing the effect of a treatment on a long-term outcome \(Y\), such as overall survival time. Because long-term outcomes may not be observable within the timeframe of a study, many researchers rely on short-term experiments to learn the treatment's effects on various short-term or "surrogate" outcomes \(X\) that predict \(Y\), such as progression-free survival and other biomarkers \citep{prentice1989surrogate,califf2018biomarker}.

Let \(d = 0\) denote the control state and \(d = 1\) the treatment state. The short-term outcomes \(Z(0)\) in the control group would typically have the same distribution as \(X\) if units are randomly selected into the experiment. On the other hand, the short-term outcomes \(Z(1)\) in the treated group might have a different distribution from \(X\).

Since long-term outcomes are not observable during short experiments, researchers often assume transportability: that the treatment does not affect the conditional expectation of the long-term outcome given the short-term outcomes; see, e.g., \cite{prentice1989surrogate,athey2016surrogates}. Formally, \(\gamma_0(z) = E\{Y \mid X = z\} = E\{Y \mid Z(d) = z\}\) for all \(z\) in the support of \(Z\) and for each \(d\). Under this assumption, the long-term average outcome in the treatment group $d$ is \(\theta_0(d) = E[\gamma_0\{Z(d)\}]\).

Based on the observed short-term outcomes, the long-term treatment effect can be identified as the difference \(\theta_0(1) - \theta_0(0)\), providing valuable insights into the treatment's long-term benefits from short-term data.  
\end{example}

\begin{example}[Average treatment effect over a target population]
Another example from causal analysis involves estimating the average potential outcome in target data using conditional means from training data. Let $X = (D, X_{2})$, where $D$ is a discrete treatment and $X_{2}$ are covariates, and let $Y = Y(D)$, where each potential outcome $Y(d)$ is independent of $D$ given $X_{2}$ in the training data. Let $Z$ be a random vector in the target data with the same dimensions as $X_{2}$ but possibly a different distribution. The parameter of interest is
\begin{equation*}
    \theta_{0} = E\{\gamma_{0}(1, Z) - \gamma_{0}(0, Z)\}. \label{param}
\end{equation*}
Under unconfoundedness, namely when the treatment is independent of the potential outcomes given $X_{2}$, $\theta_{0}$ represents the average treatment effect on the potential outcome $Y(d)$ for a target population with covariates $Z$ shifted relative to $X_{2}$ in the training population. Here, $m(z, \gamma) = \gamma(1, z) - \gamma(0, z)$. In Section~\ref{sec:application} below, we consider an empirical application that falls within this framework, where $Z$'s are characteristics of the treated units, so that the target parameter is the average effect on the treated.
\end{example}

\begin{example}[Average incremental effects over a target population]
To illustrate the generality of our approach, consider the previous setting with unconfoundedness, but now suppose $D$ is a count or continuous variable, such as education level or tax percentage. We are interested in average causal effects due to incremental changes in $d$, computed over subpopulations with characteristics $Z$:
\[
\theta_0 = \int E\{ \gamma_0(d+1, Z) - \gamma_0(d, Z)\} \, d\mu(d); \quad \theta_0 = \int E\{ \partial_D \gamma_0(d, Z)\} \, d\mu(d),
\]
averaged over values of $d$ according to a prespecified measure $\mu$. These parameters identify the average incremental treatment effect on the potential outcome $Y(d)$ for a target population with covariates $Z$. Here, $m(z, \gamma) = \int \{\gamma(d+1, z) - \gamma(d, z)\} \, d\mu(d)$ and $m(z, \gamma) = \int \{\partial_D \gamma(d, z)\} \, d\mu(d)$. For other examples that fall within this framework, see, e.g., \cite{chernozhukov2022automatic}.
\end{example}

To explain our estimation approach, we first discuss the sampling framework. Typically, $Z$ and $(Y, X)$ come from distinct, non-overlapping datasets. We consider a regression learner $\hat{\gamma}(X)$ computed from training data with $T$ observations:
$$
(Y_{1}, X_{1}), \ldots, (Y_{T}, X_{T}).
$$
Estimators of the parameter of interest also utilize $N$ observations from the target data:
$$
Z_{1}, \ldots, Z_{N}.
$$

A standard plug-in estimator of the parameter $\theta_0$ is
$$
\tilde{\theta} = \frac{1}{N} \sum_{i=1}^{N} m(Z_{i}, \hat{\gamma}).
$$
This estimator is known to suffer from bias resulting from regularization and model selection in $\hat{\gamma}$, as discussed in detail, e.g., in \citet{Chernozukov2018}. This motivates a debiased approach to eliminate or reduce the bias in $\tilde{\theta}$ arising from $\hat{\gamma}$.

The debiased estimator is obtained by adding an estimate of the debiasing term, depending on the training data:
\begin{equation*}\label{eq: bias corr}
\tilde{\theta} + \frac{1}{T} \sum_{t=1}^T \hat \alpha(X_t)\{Y_t - \hat{\gamma}(X_t)\},
\end{equation*}
where $\hat \alpha$ is an estimator of $\alpha_0$,  a balancing weight (generally, a Riesz representer) such that
\begin{equation}
E\{m(Z, \Delta)\} = E\{\alpha_{0}(X) \Delta(X)\}, \label{eq:debias_func}
\end{equation}
for all $\Delta$ with $E\{\Delta^2(X)\} < \infty$ and $E\{\alpha_{0}^2(X)\} < \infty$. By the Riesz representation theorem, the existence of $\alpha_{0}$ is equivalent to mean square continuity of $E\{m(Z, \Delta)\}$ in $\Delta$, meaning there is a constant $C<\infty$ such that $\left| E\{m(Z, \Delta)\} \right|^{2} \leq C E\{\Delta^{2}(X)\}$ for all $\Delta$ with finite second moment.

To see how debiasing works, note that for any fixed  functions $\gamma$ and $\alpha$, and $\Delta = \gamma- \gamma_{0}$, we have
\begin{align*}
     E\{m(Z,\gamma)\}-\theta_{0}  - E[\alpha(X)\{Y-\gamma(X)\}]  &  =E\{m(Z,\Delta
    )\}-E\{\alpha(X)\Delta(X)\} \\
      =E\{\alpha_{0}(X)\Delta(X)\}-E\{\alpha(X)\Delta(X)\} 
      &=-E[\{\alpha%
    (X)-\alpha_0(X)\}\{\gamma(X)-\gamma_0(X)\}],
\end{align*}
where the first equality follows from the definition of $\Delta$ and linearity, and the second equality uses equation (\ref{eq:debias_func}). Thus, the debiased expression differs from $\theta_{0}$ only by the expected product of the errors in the representer and the regression function. In particular, when $\alpha = \alpha_{0}$ or $\gamma = \gamma_{0}$, the bias vanishes completely. This is known as a double robustness property, as demonstrated by \citet{chernozhukov2018global} for the general case of target parameters given as linear functionals of a regression.

Using this bias correction to estimate the parameter of interest depends crucially on being able to estimate $\alpha_{0}(X)$ from equation (\ref{eq:debias_func}). This $\alpha_{0}$ can be identified as the minimizer of the following variational problem:
\begin{equation}
\alpha_{0} = \arg\min_{\alpha \in \mathcal{A}} \left[ E\{\alpha^{2}(X)\} - 2 E\{m(Z, \alpha)\} \right], \label{eq:alpha0}
\end{equation}
where $\mathcal{A}$ is the parameter space containing the true value $\alpha_{0}$. Indeed, by adding and subtracting $E\{\alpha_{0}^{2}(X)\}$ and completing the square, we obtain
\begin{align*}
E\{-2 m(Z, \alpha) + \alpha^{2}(X)\} &= - E\{\alpha_{0}^{2}(X)\} + E\{\alpha_{0}^{2}(X) - 2 \alpha_{0}(X) \alpha(X) + \alpha^{2}(X)\} \\
&= - E\{\alpha_{0}^{2}(X)\} + E[\{\alpha_{0}(X) - \alpha(X)\}^{2}],
\end{align*}
which is minimized at $\alpha(X) = \alpha_{0}(X)$. Consequently, an estimator of $\alpha_{0}$ can be constructed by minimizing a sample version of the objective function in equation (\ref{eq:alpha0}), using data on $X$ and $Z$ to estimate the first and second components of the loss function. We describe formal procedures in the next subsection. The justification of equation (\ref{eq:alpha0}) is similar to that in \citet{chernozhukov2021automatic}, with the difference that here $Z$ comes from the target population, while $Y$ and $X$ from the training population.
\subsection{Debiased estimation with cross-fitting}
\label{section:crossfit}

We shall use cross-fitting, a form of sample splitting, in the construction of the debiased machine learner. Cross-fitting is known to reduce bias for some
estimators further and to help obtain large sample inference results for a rich variety of
regression learners under minimal conditions.  

Let $I_{\ell},(\ell=1,\ldots,L)$ denote a partition
of the training sample indices $\{1, \ldots, T\}$ into $L$ distinct subsets of about equal
size. Let $T_{\ell}$ be the number of observations in $I_{\ell}.$ In
practice, $L=5$ (5-fold) or $L=10$ (10-fold) cross-fitting is often used. Also,
let $\hat{\gamma}_{\ell}(x)$ and $\hat{\alpha}_{\ell}(x)$ be
estimators of $\gamma_{0}$ and $\alpha_{0}$, respectively, computed from all observations not
in $I_{\ell}$. We give details below.

The regression learner is the solution to the problem
$$
\begin{gathered}
\min_{\gamma \in \mathcal{G}} \frac{1}{T-T_\ell}\sum_{t \notin I_{\ell}}\left\{Y_t-\gamma\left(X_t\right)\right\}^2+\operatorname{pen}(\gamma), 
\end{gathered}
$$
where $\gamma$ is a function in a set $\mathcal{G}$ and $\operatorname{pen}(\gamma)$ is a penalty function. For example, in the case of lasso,
$$\gamma(x)=b\left( x\right)^{\T} \beta; \quad \operatorname{pen}(g)=r_\gamma \sum_{j=1}^p\left|\beta_j\right|
$$
where $b\left(x\right) = \{b_1(x), \ldots, b_p(x)\}^\T$ is a dictionary of approximating functions, for example polynomials and interactions; $\beta$ is a vector of parameters defining $\gamma$; and $r_\gamma$ is the penalty level.

The Riesz learner is the solution to the ``Riesz regression" problem
$$
\begin{aligned}
& \min_{\alpha \in \mathcal{A}} \frac{1}{T-T_\ell} \sum_{t \not \in I_{\ell}} \alpha^2\left(X_t\right)- 2 \frac{1}{N} \sum_{i=1}^N m\left(Z_i, \alpha \right)+\operatorname{pen}(\alpha),
\end{aligned}$$
where $\alpha$ is a function in a set $\mathcal{A}$ and $\operatorname{pen}(\alpha)$ is a penalty function.  Here, $\hat{\alpha}_{\ell}$ minimizes an objective function where the the quadratic term is obtained from the training data and the linear term
from the target data. For example, for the lasso learner, we set
$$
\begin{aligned}
\alpha\left(x\right)=b\left(x \right)^{\T} \rho ; \quad  \operatorname{pen}(a)=r_a \sum_{j=1}^p\left|\rho_j\right|,
\end{aligned}
$$
where $\rho$ are the parameters defining $\alpha$, and $r_a$ is the penalty level. It differs from that of
\citet{chernozhukov2022automatic} in the quadratic and linear terms coming
from different data sets. 

For each fold $\ell\,\ $of the cross-fitting, a debiased machine learner can be constructed as the sum of a plug-in estimator and a bias correction
\begin{equation*}
    \hat{\theta}_{\ell}=\frac{1}{N}\sum_{i=1}^{N}m(Z_{i},\hat{\gamma}_{\ell
    })+\frac{1}{T_{\ell}}\sum_{t\in I_{\ell}}\hat{\alpha}_{\ell}(X_{t})\{Y_{t}
    -\hat{\gamma}_{\ell}(X_{t})\}. \label{eq:theta_hat}
\end{equation*}
To estimate the asymptotic variance for each $\hat\theta_{\ell}$, we trim the estimator of the debiasing function to obtain $\tilde{\alpha}_{\ell}(X)=\tau_N\{\hat{\alpha}_{\ell}(X)\}$, where
\begin{equation*}
\tau_N(a) = \min(|a|,\bar{\tau}_N) \rm{sign}(a) 
\end{equation*}
and $\bar{\tau}_N$ is a large positive constant that grows slowly with $N$. 
The purpose of this trimming is guarantee consistency of the asymptotic variance estimator when we only have mean square convergence rates for the estimators of the regression and the debiasing function.
The asymptotic variance of $N^{1/2}(\hat\theta_{\ell}-\theta_0)$ can be estimated as
\begin{align*}
    \hat{V}_{\ell}  & =\hat{s}_{\ell m}^{2}+\frac{N}{T}\hat{s}_{\ell\alpha}^{2},\text{ } \\
    \hat{s}_{\ell m}^{2} & =\frac{1}{N}\sum_{i=1}^{N}\{m(Z_{i},\hat{\gamma}_{\ell
    })-\bar{m}_{\ell}\}^{2}, \quad 
    \bar{m}_{\ell}  =\frac{1}{N}\sum_{i=1}^{N}m(Z_{i},\hat{\gamma}_{\ell})\\ \text{ }\hat{s}_{\ell\alpha}^{2}&=\frac{1}{T_{\ell}%
    }\sum_{t\in I_{\ell}}\tilde{\alpha}_{\ell}(X_{t})^{2}\{Y_{t}-\hat{\gamma}_{\ell
    }(X_{t})\}^{2}.
\end{align*}
A single bias-corrected estimator and asymptotic variance estimator can then
be obtained by a weighted average of the estimators across the sample splits:
$$
\hat{\theta}=\sum_{\ell=1}^{L}\frac{T_{\ell}}{T}\hat{\theta}_{\ell},\quad \hat{V}=\sum_{\ell=1}^{L}\frac{T_{\ell}}{T}\hat{V}_{\ell}.
$$

\subsection{Large sample theory with cross-fitting}

We show asymptotic normality of the cross-fit estimator under weak regularity conditions. Let $\left\Vert a\right\Vert _{2}=[E\{a^2(X)\}]^{1/2}$ be the mean square norm of a measurable function $a(X)$. Here and below, all quantities are implicitly indexed by $N$, and the asymptotic results are given under $N \to \infty$.

\begin{assumption}[Covariate shift parameter]\label{assumption:bounded}
    a) The defining formula $m(Z,\gamma)$ is linear in $\gamma$.
    b) The training data $(Y_{t},X_{t})$ for $(t=1,...,T)$, and the target data $(Z_{i})$ for $(i=1,...,N)$, are each generated as independent and identically distributed random vectors, with the two samples mutually independent.
    c) The ratio of target to training data sizes, $N/T$ converges to $0 \leq \xi<\infty$.
    d)  The functional $m(Z,\gamma)$ is mean square continuous in $\gamma$, namely $E\{m(Z,\gamma)^{2}\}\leq C\left\Vert \gamma\right\Vert _{2}^{2}$ for some $C<\infty$.
    e) The representer $\alpha_{0}(X)$ and conditional variance  $\text{var}(Y|X)$ are bounded.
\end{assumption}

This condition differs from the previous work on Riesz estimation in imposing combined assumptions on both training and target data.  The last condition simplifes the analysis and presentation, but can be replaced by more general conditions.

\begin{assumption}[Consistent learners]\label{assumption:learners}
    For each $\ell\in L$, the learners $\hat{\gamma}_{\ell}$ and $\hat{\alpha}_{\ell}$ are mean square consistent: $\Vert\hat{\gamma}_{\ell}-\gamma_{0}\Vert_{2}=o_{p}(1)$, and
  $  \Vert\hat{\alpha}_{\ell}-\alpha_{0}\Vert_{2}=o_{p}(1).$ The product of their convergence rates is faster than $N^{-1/2}$: $\Vert\hat{\alpha}_{\ell}-\alpha_{0}\Vert_{2}
\Vert\hat{\gamma}_{\ell}-\gamma_{0}\Vert_{2}
    =o_{p}(N^{-1/2}).$
\end{assumption}

This condition requires consistency of regression and representer learners, while allowing for a tradeoff in the accuracy of estimating the regression and the representer. Under these conditions, we prove quasi-oracle properties for the debiased estimator.
\begin{lemma}[Quasi-oracle estimator]\label{lemma:cross}
If Assumptions~\ref{assumption:bounded} and~\ref{assumption:learners} are satisfied, the prelimary estimation of $\gamma_0$ and $\alpha_0$ does not affect the first-order large-sample properties of the estimator:
$$
\hat{\theta}=\frac{1}{N}\sum_{i=1}^{N}m(Z_{i},\gamma_{0})+\frac
{1}{T}\sum_{t=1}^{T}\alpha_{0}(X_{t})\{Y_{t}-\gamma_{0}(X_{t})\}+o_{p}%
(1/N^{1/2}).
$$
\end{lemma}

This implies $N^{-1/2}$ consistency and asymptotic normality of the cross-fitted debiased estimator.

\begin{theorem}[Asymptotically normal estimator]\label{theorem:cross}
    If Assumptions~\ref{assumption:bounded} and~\ref{assumption:learners} are satisfied, then $N^{1/2}(\hat\theta-\theta_0)$ converges in distribution to a centered Gaussian variable $N(0,V)$, with variance $V={\normalfont \text{var}}\{m(Z,\gamma_0)\} + {\xi}E\{\alpha_0^2(X){\normalfont \text{var}}(Y|X)\}$. Also, if  $\bar{\tau}_N\Vert\hat{\gamma}_{\ell}-\gamma_{0}\Vert_{2} =o_{p}(1)$ for each $\ell$, then the variance estimator is consistent: $\hat{V} = V + o_{p}(1)$.
\end{theorem}
The asymptotic variance depends on the ratio of sample sizes $\xi = \lim N/T \geq 0$. The variance increases when the training sample size is small relative to the target sample size, and vice versa.



\subsection{Debiased estimation without cross-fitting via lasso}

Cross-fitting enables high-quality inference with very few assumptions.  When we can adequently control the complexity of the machine learners of $\gamma_0$ and $\alpha_0$ to avoid overfitting biases, we don not need to rely on cross-fitting; see, e.g., \cite{belloni2015uniform,belloni2017program}. We highlight this possibility in the data-fused setting using lasso-type learners with carefully chosen penalty levels.

To describe the full-sample estimator, let $b(x)$ be the $p\times1$
dictionary of basis functions. The lasso regression estimator
from the training data is
\begin{equation}
    \hat{\gamma}(x)=b(x)^{\T}\hat{\beta}, \quad \hat \beta =\arg\min_{\beta}\frac{1}{T}\sum
    _{t=1}^{T}\{Y_{t}-b(X_{t})^{\T}\beta\}^{2}+2r_{\gamma}\sum_{j=1}^{p}\left\vert
    \beta_{j}\right\vert .
    \label{eq:lasso}
\end{equation}
A corresponding $\hat{\alpha}$  takes the form
\begin{equation}\hat{\alpha
}(x)=b(x)^{\T}\hat{\rho}\quad \hat{\rho}=\arg\min_{\rho}\{-2\hat{M}^{\T}\rho+\rho^{\T}\hat{Q}%
    \rho+2r_{\alpha}\sum_{j=1}^{p}\lvert\rho_{j}\rvert\},
    \label{eq:rhohat}
\end{equation}
where 
$$
\hat{M}_{j}=\frac{1}{N}\sum_{i=1}^{N}m(Z_{i},b_{j}),\quad \hat{M}=(\hat
{M}_{1},\ldots,\hat{M}_{p})^{\T},\quad \hat{Q}=\frac{1}{T}\sum_{t=1}%
^{T}b(X_{t})b(X_{t})^{\T}.
$$

A debiased machine learner without cross-fitting is%
\begin{align*}
    \hat{\theta}  &  =\frac{1}{N}\sum_{i=1}^{N}m(Z_{i},\hat{\gamma})+\frac{1}{T}\sum_{t=1}^{T}\hat{\alpha}(X_{t})\{Y_{t}-\hat{\gamma}(X_{t})\}.
     \label{eq:deb_noxfit}
\end{align*}
Equations \eqref{eq:lasso}--\eqref{eq:rhohat} show that the only features of the training data needed to construct this estimator are the second moment matrix $\hat{Q}$ of the
dictionary and the cross product $\sum_{t=1}^{T}b(X_{t})\left\{  Y_{t}%
-\hat{\gamma}(X_{t})\right\}  $ between the observations on the dictionary  and the lasso residuals.

An asymptotic variance estimator without cross-fitting is then
$$
\hat{V}=\frac{1}{N}\sum_{i=1}^{N}\{m(Z_{i},\hat{\gamma})-\bar{m}\}^{2}+\frac{N}{T}\frac{1}%
{T}\sum_{t=1}^{T}\tilde{\alpha}(X_{t})^{2}\{Y_{t}-\hat{\gamma}(X_{t})\}^{2},\quad \bar{m}=\frac{1}{N}\sum_{i=1}^{N}m(Z_{i},\hat{\gamma}),
$$
where $\tilde{\alpha}(X_{t})=\tau_N\{\hat{\alpha}(X_{t})\}$.

\subsection{Large sample theory without cross-fitting}
Next we give conditions that will be sufficient for asymptotic normality and consistent asymptotic variance estimation when lasso is used on the training data without cross-fitting.
\begin{assumption}[Approximately sparse regression]\label{assumption:sparse}
    The projection of the regression onto the dictionary is approximately sparse: there exist
    $C<\infty$ and $\zeta>1/2$ such that
    for all $\beta$ minimizing $E[\{\gamma_0(X)-b(X)^\T\beta\}^2]$ and for all sparsity levels $s\leq p$, and
    there is some vector $\tilde{\beta}$ with $s$ nonzero elements and $\Vert\beta-\tilde{\beta}\Vert\leq Ct^{-\zeta}.$
\end{assumption}
This condition requires that the sparse approximation rate for the regression must be greater than 1/2.  Let $p$ denote the dimension of the dictionary $b=b^p$. Let $\mathcal{B}^p$ denote the mean square closure of the linear span of $b^p(X)$, and $\mathcal{B} = \cup_{p=1}^\infty\cap^\infty_{m=p} \mathcal{B}^m$ be its limit as $p \to \infty$. Here, $\mathcal{B}$ is the space of regression functions that is eventually approximable by the sequence of dictionaries $(b^p)$.

\begin{assumption}[Covariate shift parameter]\label{assumption:regular}
    The defining formula satisfies Riesz representation within $\mathcal{B}$: there is some $\alpha_0\in\mathcal{B}$ such that $E\{m(Z,\gamma)\}=E\{\alpha_0(X)\gamma(X)\}$ for all $\gamma\in\mathcal{B}$. The defining formula applied to the dictionary is bounded:
    $m(Z,b_{j})=\tilde{m}_{0}(Z)\tilde{m}_{j}(Z)$, where $E\{\tilde{m}
    _{0}(Z)^{2}\}<\infty$ and $\max_{j\leq p}\left\vert \tilde{m}_{j}(Z)\right\vert \leq C$ for some $C<\infty$. 
    
\end{assumption}
This condition ensures that sample averages of $m(Z_i,b_{j})$ converge to their expectations uniformly in $j\leq p$ at the familiar rate $\varepsilon_N=\{\ln(p)/N\}^{1/2}$.  

\begin{assumption}[Regularization]\label{assumption:regularization} The penalty levels $r=r_\gamma$ and $r=r_\alpha$ are slightly slower than $\varepsilon_N$:
    $\varepsilon_{N}=o(r)$ and $r=o(N^{c} \varepsilon_{N})$ for every $c>0.$
\end{assumption}

By allowing penalty levels to vanish slightly slower than $\varepsilon_{N}$, we simplify the statement of conditions. This assumption could be avoided by specifying that $r\approx C\varepsilon_{N}$ for a large enough $C$, at the cost of more complicated exposition. 
\begin{assumption}[Dictionary regularity]\label{assumption:dictionary} The elements of the dictionary are uniformly bounded and the population Gram matrix $Q=E\{b(X)b(X)^\T\}$ is uniformly bounded: for some $C<\infty$  $\max_{j \leq p }|b_j(X)|\leq C $
    and  $\max_{\|a\|=1}a^\T Qa \leq C$, uniformly in $p$.
\end{assumption}

This condition simplifies the analysis considerably. 
 Let $J$ denote a subvector of $(1,...,p)$, $\left\vert J\right\vert $ denote
the number of elements of $J,$ $\beta_{J}$ be the vector consisting of
$\beta_{Jj}=\beta_{j}$ for $j\in J$ and $\beta_{Jj}=0$ otherwise, and
$\beta_{J^{c}}$ be the corresponding vector for $J^{c}$. Hence 
$\beta=\beta_{J}+\beta_{J^{c}}$.

\begin{assumption}[Restricted eigenvalue]\label{assumption:eigenvalue} Sparse coefficients have a minimal eigenvalue: there is $c>0$, such that given any degree of sparsity $s=O\{n/\ln(p)\}$, with probability $1- o(1)$, the sparse restricted eigenvalue is bounded: $
    \min_{\left\vert J\right\vert \leq s}\min_{\left\Vert \beta_{J^{c}%
    }\right\Vert _{1}\leq3\left\Vert \beta_{J}\right\Vert _{1}}
    {\beta^{\T}\hat{Q}\beta       }/{\beta_{J}^{\T}\beta_{J}}\geq c.
    $
\end{assumption}
This is a standard restricted eigenvalue condition on the Gram matrix $\hat Q$ for lasso and related estimators \citep{bickel2009simultaneous}. 

The following assumption states that $\gamma_0$ can be approximated by $b$ as $p$ increases.

\begin{assumption}[Correct specification]\label{assumption:correct} The true regression function $\gamma_0(X) = E(Y \mid X)$ is in $\mathcal{B}$. For some $c > 0$,
 $E(|Y|^{2+c})<\infty.$ 
\end{assumption}

For the next condition, let $\gamma_N(X)$ and $\alpha_N(X)$ denote the least squares projections of $\gamma_0(X)$ and $\alpha_0(X)$, respectively, on $b(X)$. Also let $\delta_{N}=\left\Vert \alpha_0-\alpha_{N}\right\Vert_{2}$.

\begin{assumption}[Small projection errors]\label{assumption:approx}
     The regularization is small enough: $\max (r_\gamma, r_\alpha)=o(\varepsilon_{N}/\delta_N)$. The product of projection errors is small:
    $N^{1/2}\left\Vert \gamma_{0}-\gamma_{N}\right\Vert_{2}
    \left\Vert \alpha_0-\alpha_{N}\right\Vert _{2}
 =o(1)$.
\end{assumption}


The former condition limits the extent to which penalty levels can be larger than  $\varepsilon_N$, highlighting the delicacy of choosing penalty levels without cross-fitting.

The following result follows similarly to Corollary 9 of~\cite{bradic2022minimax}. 

\begin{theorem}[Asymptotically normal estimator]\label{theorem:lasso}
    If Assumptions~\ref{assumption:bounded},~\ref{assumption:sparse},~\ref{assumption:regular},~\ref{assumption:regularization},~\ref{assumption:dictionary},~\ref{assumption:eigenvalue},~\ref{assumption:correct}, and~\ref{assumption:approx} are satisfied then $N^{1/2}(\hat\theta-\theta_0)$ converges in distribution to a centered Gaussian variable $N(0,V)$, with variance $V={\normalfont\text{var}}\{m(Z,\gamma_0)\} + {\xi}E\{\alpha_0(X)^2{\normalfont\text{var}}(Y|X)\}$.
    If, in addition, for some $c>0$, the growth condition $\bar{\tau}_{N}\{N^c\varepsilon_N^{(2\zeta-1)/(2\zeta+1)}+\left\Vert \gamma_{N}-\gamma_{0}\right\Vert_{2}\} =o(1)$ is satisfied, then the variance estimator is consistent: $\hat{V} = V + o_{p}(1)$. 
\end{theorem} 

\section{Simulaton experiments}\label{sec:sim}



We evaluate the proposed method through simulated experiments, setting $\gamma_0$ as a high-dimensional, high-order polynomial defined by
\begin{equation*}
\gamma_0(x) = \sum_{q=1}^Q \delta_q \left( \sum_{k=1}^K \iota_{qk} x_k \right)^q.
\end{equation*}
 This polynomial is of order $Q=3$ over $K=6$-dimensional vectors $x$, includes cross terms, and is parameterized by $\delta$ and  $\iota$. The outcome is generated as $
Y_t = \gamma_0(X_t) + \epsilon_t,$
where $\epsilon_t \sim N(0, \sigma^2)$ with $\sigma = 0.1$.
The components of $X_t$ are independently generated from a mixture of $N(0,1)$ and $U(-5,5)$ distributions with mixing probabilities $0.99$ and $0.01$, respectively. The shifted covariates $Z$ are generated from $N(\mu, 1)$ with a mean shift $\mu = 1.1 \sigma$.  The $\delta_q$ parameters  are set as $\delta_1 = 1$, $\delta_2 = 0.7$, and $\delta_3 = 0.5$. 

Multiple simulation specifications are created by randomizing the parameters $\iota_k=(\iota_{1k}, \iota_{2k},\iota_{3k})^T$. The parameters $\iota_{jk}$ are drawn from $U(-R_k, R_k)$, where $R_1 = 1$, $R_2 =R_3=R_4 = R_5 = 0.8$, and $R_6 = 0.16$. To induce sparsity, 60\% of these draws are set to zero.  We vary the simulation specifications and the training dataset initialization parameters to compute performance statistics. Each simulation uses $10,000$ observations of $X$ and $Z$. We generate $27$ specifications and draw samples from each $60$ times, resulting in $1,620$ experiments.

\begin{figure*}[ht]
	\centering \begin{subfigure}[c]{0.49\textwidth}
\includegraphics[trim={4.3cm 8cm 4.3cm  8cm}, width=\textwidth, clip]   {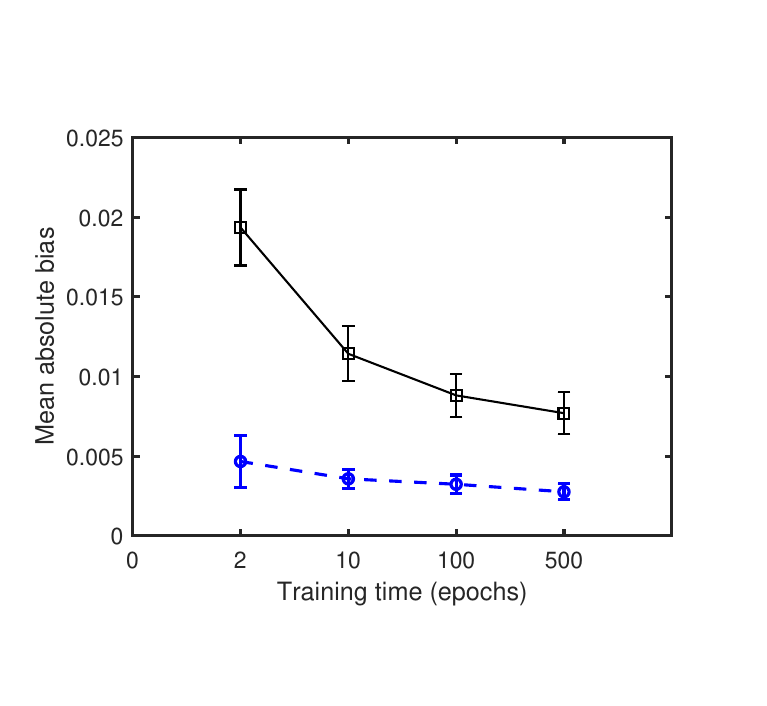}
\caption{Average absolute bias}
	\end{subfigure}
    \begin{subfigure}[c]{0.49\textwidth}
\includegraphics[trim={4.3cm 8cm 4.3cm  8cm}, width=\textwidth, clip]  {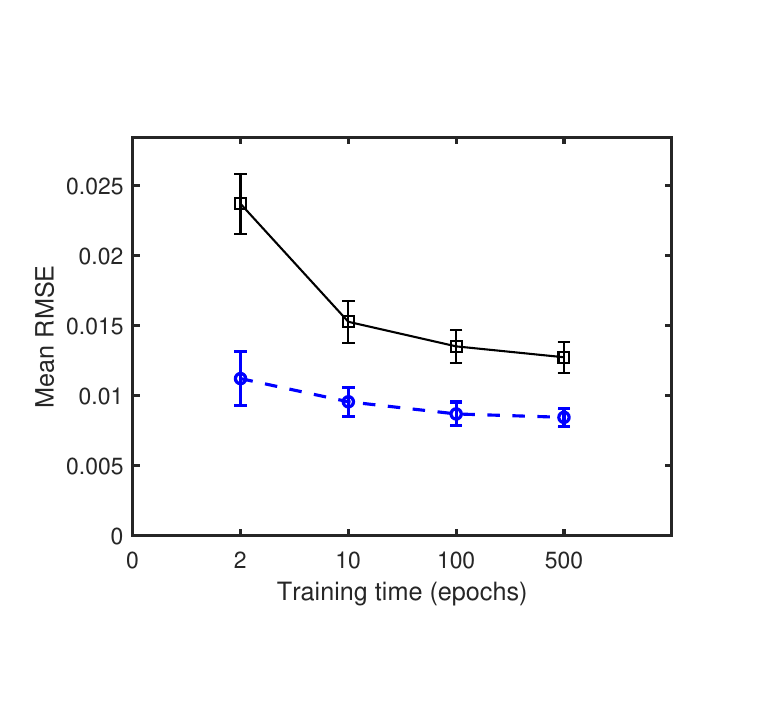}
\caption{Root mean square error (RMSE)}
	\end{subfigure}
    \hfil
    \caption{    
    We compare the plug-in estimator (solid line) and our estimator (dashed line), with error bars (vertical intervals). The error bars give standard deviation across different specifications.}
\label{fig:bias_comb}
\end{figure*}

We construct the regression estimator $\hat{\gamma}$ using a standard fully connected neural network. We conduct estimation with $2$, $10$, $100$, or $500$ epochs of training. We interpret different numbers of epochs as as different training times for the neural network. The Supplementary Material gives implementation details.

We estimate the lasso Riesz representer with a dictionary vector $b(x)$ consisting of $p$ quadratic basis functions: $b(x) = (1, x_1, \ldots, x_K, x_1 x_1, \ldots, x_1 x_K, \ldots,  x_K x_K)^T.$  We evaluate the algorithm's effectiveness by computing the average absolute bias and the root mean square error (RMSE) in estimating the target parameter $\theta_0 = E\{\gamma_0(Z)\}$. We compare the performance of the plug-in estimator with that of the debiased estimator.



Figure~\ref{fig:bias_comb} plots the average absolute bias and average RMSE as functions of the neural network epochs, interpretable as the neural network training times. The bias correction consistently reduces the average bias and RMSE across specifications, with the largest improvement at shorter training times. This can be explained by increased bias due to regularization, since lower training times correspond to early stopping regularization.  
Figure~\ref{fig:bias_comb} also includes error bars showing the estimated standard deviations of the mean absolute bias estimates and mean RMSEs. The bias correction is strongly beneficial across specifications. Furthermore, the bias correction achieves nearly the same RMSE at 100 epochs as at 500 epochs, allowing practitioners to reduce training times considerably.

\section{Difference-in-differences estimation of minimum wage effect on teen employment}\label{sec:application}

We apply our methodology to estimate the average treatment effect on the treated using unconfounded difference-in-differences identification. We have data $(Y_{0,i}, Y_{1,i}, D_i, X_i)$ for $(i=1,...,n)$, which are independent and identically distributed copies of the random variables $(Y_0, Y_1, D, X)$. Here, $Y_0$ and $Y_1$ are outcomes in two time periods for the same unit, $D$ indicates whether the unit was treated at the end of the first period, and $X$ are control variables. Under the conditional parallel trends assumption (e.g., \cite{roth2023s}), the effect is identified as
\begin{align*}
\theta_0 = E\{\Delta Y - \gamma_0(X) \mid D=1\}, \quad \gamma_0(X) = E\{\Delta Y \mid D=0, X\},\quad \Delta Y = Y_1 - Y_0.
\end{align*}

This estimand is a functional of the regression function $\gamma_0$, trained using data from the untreated subgroup ($D=0$), over the distribution of data in the treated subgroup ($D=1$); there is a covariate shift. We apply our automatic debiasing approach and estimate the Riesz representer by minimizing the empirical analogue of the Riesz loss:
\begin{align*}
\min_{\alpha \in \mathcal{A}} E\{\alpha^2(X) \mid D=0\} - 2 E\{-\alpha(X) \mid D=1\}.
\end{align*}
We consider the resulting automatically debiased cross-fitted estimate of the target parameter, described in Section~\ref{section:crossfit}. 

We implement a variant of the cross-fitting approach that splits the treated and untreated subgroups into $L$ folds. While cross-fitting the treated subgroup is not required for valid inference, it helps in estimating the Riesz representer esimator's performance via the cross-fitted Riesz loss.

\begin{table}
\fontarxivtable	
	\tbl{Comparison of performance of different estimation variants on estimation of ATT of minimum wage above federal in 2004 on teen employment in years 2004 and 2005 via difference-in-differences identification. }
{
\begin{tabular}{rcccccccc}
{} & \multicolumn{4}{c}{Year $=$ 2004} & \multicolumn{4}{l}{ Year $=$ 2005 } \\
{} & RMSE & Riesz loss &    ATT &    s.e. & RMSE & Riesz loss &    ATT &    s.e. \\
No controls    &   0.163 & -1.000 & -0.040 & 0.019 &   0.188 & -1.000 & -0.076 & 0.020 \\
\multicolumn{1}{l}{{ Propensity}} \\
({LR}, {LG})          &   0.163 & -3.077 & -0.024 & 0.020 &   0.185 & -2.883 & -0.047 & 0.021 \\
({LR-int}, {LG-int})      &   0.164 & -3.026 & -0.022 & 0.020 &   0.186 & -2.797 & -0.048 & 0.021 \\
({Lasso-3d}, {LG-3d})     &   0.163 & -1.462 & -0.027 & 0.019 &   0.185 & -1.909 & -0.044 & 0.020 \\
({Ridge-3d}, {LG-3d})     &   0.163 & -1.462 & -0.026 & 0.019 &   0.185 & -1.909 & -0.044 & 0.020 \\
({RF}, {RF})  &   0.165 & -3.035 & -0.023 & 0.019 &   0.188 & -2.854 & -0.049 & 0.020 \\
\multicolumn{1}{l}{{ AutoDML}}\\
({LR-int}, {Lasso-RR-int})        &   0.164 & { -3.527} & -0.024 & 0.020 &   0.186 & { -3.439} & -0.043 & 0.021 \\
({RF}, {RF-RR})   &   0.165 & -2.929 & -0.022 & 0.019 &   0.188 & -2.854 & -0.048 & 0.020 \\
({Lasso-3d}, {Lasso-RR-3d}) &   0.163 & -3.395 & -0.024 & 0.020 &   0.185 & -3.140 & -0.045 & 0.021 \\
({RF}, {NN-RR})   &   0.165 & -2.783 & -0.022 & 0.019 &   0.188 & -2.799 & -0.046 & 0.020 
\end{tabular}
}
	\label{table:1}
	\begin{tabnote}
	ATT, estimate of average treatment effect on the treated; s.e., standard error of ATT estimate; RMSE, out-of-sample root mean squared error of $\hat \gamma$ in predicting $\Delta Y$; Riesz loss, out-of-sample Riesz loss of $\hat{\alpha}$ in estimating $\alpha_0$.
	\end{tabnote}
\end{table}

\begin{table}
\fontarxivtable
	\tbl{Comparison of performance of different estimation variants on estimation of ATT of minimum wage above federal in 2004 on teen employment in years 2006 and 2007 via difference-in-differences identification.}
{
\begin{tabular}{rcccccccc}
{} & \multicolumn{4}{l}{Year $=$ 2006} & \multicolumn{4}{l}{Year $=$ 2007} \\
{} & RMSE & Riesz loss &    ATT &    s.e. & RMSE & Riesz loss &    ATT &    s.e. \\
{No controls}    &   0.223 & -1.000 & -0.117 & 0.020 &   0.230 & -1.000 & -0.131 & 0.023 \\
\multicolumn{1}{l}{{ Propensity}} \\
({LR}, {LG})         &   0.217 & -3.064 & -0.053 & 0.020 &   0.222 & -5.538 & -0.067 & 0.023 \\
({LR-int}, {LG-int})      &   0.218 & -2.899 & -0.053 & 0.020 &   0.222 & -5.273 & -0.066 & 0.024 \\
({Lasso-3d}, {LG-3d})      &   0.217 & -1.535 & -0.057 & 0.020 &   0.222 & -1.324 & -0.068 & 0.022 \\
({Ridge-3d}, {LG-3d})     &   0.218 & -1.535 & -0.055 & 0.020 &   0.222 & -1.324 & -0.068 & 0.022 \\
({RF}, {RF})  &   0.220 & -3.253 & -0.051 & 0.020 &   0.223 & -4.968 & -0.067 & 0.023 \\
\multicolumn{1}{l}{{ AutoDML}}\\
({LR-int}, {Lasso-RR-int})        &   0.218 & { -3.483} & -0.051 & 0.020 &   0.222 & { -6.014} & -0.065 & 0.025 \\
({RF}, {RF-RR})  &   0.220 & -3.220 & -0.055 & 0.020 &   0.223 & -4.649 & -0.064 & 0.023 \\
({Lasso-3d}, {Lasso-RR-3d}) &   0.217 & -3.067 & -0.052 & 0.021 &   0.222 & -5.989 & -0.060 & 0.025 \\
({RF}, {NN-RR})   &   0.220 & -2.892 & -0.052 & 0.020 &   0.223 & -4.660 & -0.064 & 0.023 
\end{tabular}
}
	\label{table:2}
	\begin{tabnote}
	ATT, estimate of average treatment effect on the treated; s.e., standard error of ATT estimate; RMSE, out-of-sample root mean squared error of $\hat \gamma$ in predicting $\Delta Y$; Riesz loss, out-of-sample Riesz loss of $\hat{\alpha}$ in estimating $\alpha_0$.
	\end{tabnote}
	\label{tab2}
\end{table}

\begin{table}\tbl{Pre-trend test for violations of difference-in-differences identification.}
\fontarxivtable
{\begin{tabular}{rcccc}
& \multicolumn{4}{l}{Year $=$ 2002} \\
{} & RMSE & Riesz loss &    ATT &    s.e. \\
No controls    &   0.154 & -1.000 & -0.005 & 0.013 \\
\multicolumn{1}{l}{{ Propensity}} \\
({LR}, {LG})          &   0.154 & -2.639 &  0.003 & 0.014 \\
({LR-int}, {LG-int})      &   0.154 & -2.555 &  0.002 & 0.014 \\
({Lasso-3d}, {LG-3d})     &   0.155 & -1.849 & -0.003 & 0.013 \\
({Ridge-3d}, {LG-3d})     &   0.154 & -1.849 & -0.001 & 0.013 \\
({RF}, {RF})  &   0.155 & -1.546 &  0.002 & 0.013 \\
\multicolumn{1}{l}{{ AutoDML}}\\
({LR-int}, {Lasso-RR-int})        &   0.154 & { -3.032} &  0.005 & 0.014 \\
({RF}, {RF-RR})   &   0.155 & -2.606 &  0.000 & 0.012 \\
({Lasso-3d}, {Lasso-RR-3d}) &   0.155 & -2.790 &  0.000 & 0.014 \\
(RF, NN-RR)   &   0.155 & -2.600 &  0.001 & 0.012 
\end{tabular}
}
	\label{table:3}
	\begin{tabnote}
	ATT, estimate of average treatment effect on the treated; s.e., standard error of ATT estimate; RMSE, out-of-sample root mean squared error of $\hat \gamma$ in predicting $\Delta Y$; Riesz loss, out-of-sample Riesz loss of $\hat{\alpha}$ in estimating $\alpha_0$.
	\end{tabnote}
\end{table}

We estimate the effect of minimum wage changes on teen employment using data from \citet{callaway2021difference}. The data consist of annual county-level records from the United States covering 2001 to 2007. The outcome variable is the logarithm of county-level teen employment, and the treatment variable indicates whether the county has a minimum wage above the federal minimum wage. The data also include county population and average annual pay. We assume the parallel trends assumption holds conditional upon the pre-treatment variables, which are: 2001 population, 2001 average pay, 2001 teen employment, and indicators for the four regions where the counties are located.

We estimate the effect of treatment in 2004 on all subsequent years using our automatic debiasing procedure. We employ various methods for estimating the Riesz representer $\alpha_0$ and the regression function $\gamma_0$, including linear models, neural networks, random forests, and high-dimensional regularized linear models. Tables~\ref{table:1} and~\ref{table:2} report the empirical findings. We also check for a violation of the difference-in-differences assumption by estimating the effect of treatment in 2004 on employment in 2002, which should be statistically indistinguishable from zero; the results are reported in Table~\ref{table:3}.

As a benchmark, the ``no controls'' method disregards the control variables $X$, so $\gamma_0$ and $\alpha_0$ are constant functions. 

The ``propensity'' methods use the explicit formula for the Riesz representer, involving the propensity model $\pi_0(X) = \Pr(D=1 \mid X)$. Each propensity variant is specified by the methods used to estimate the regression $\hat \gamma$ and the propensity $\hat{\pi}$, stated as a pair within parentheses. In the first propensity approach,
``LR'' and ``LG'' denote linear and logistic regression on the features $X$, respectively. ``LR-int'' and ``LG-int'' denote linear and logistic regression on $X$, with interactions of region indicators with baseline population, average pay, and teen employment in 2001. {``Lasso-3d''}, {``Ridge-3d''}, and {``LG-3d''} refer to lasso, ridge, and logistic regression, respectively, using third-degree polynomials of $X$ with cross-validated penalty levels. {``RF''} denotes random forest regression or classification.

The ``AutoDML'' methods use the automatic debiasing approach for the Riesz representer. As before, the first element of each pair is the regression estimator while the second element is the Riesz estimator. ``Lasso-RR-int'' is an $\ell_1$-penalized linear Riesz representer with cross-validated penalty, using $X$ and interactions as above. ``Lasso-RR-3d'' is similar, with third-degree polynomials of $X$.  ``RF-RR'' uses a random forest to minimize the Riesz loss.  Finally, ``NN-RR'' uses neural network to minimize the Riesz loss, using stochastic gradient descent with early stopping based on the out-of-sample Riesz loss.

Every method that controls for $X$ finds a similar effect. Compared to the benchmark that disregards $X$, the effect is smaller in absolute value and has a similar standard error. 

When it comes to estimating the Riesz representer, our automatic debiased machine learning approach outperforms the alternative approaches based on explicit propensity score estimation. In particular, the smallest out-of-sample Riesz loss is achieved by an ``AutoDML'' method, namely ``Lasso-RR-int''. A smaller Riesz loss implies that the estimated Riesz representer performs better covariate balancing, leading to a more robust estimate and confidence interval for the effect. 


\section{Discussion}In this paper, we provided debiased machine learning estimators for a broad class of regression functionals under covariate shift; these parameters arise in various predictive and causal inference problems. One example is estimating the long-term average effect of a treatment using short-term experimental data, where the treatment changes the distribution of predictors for long-term outcomes and separate data link these predictors to long-term outcomes. Another important example involves potential outcomes induced by a policy that shifts covariate distributions; for instance, we may be interested in changes in average house values after a policy that improves the environmental characteristics of a neighborhood.

We developed a method to debias estimators of these functionals using auxiliary Riesz regressions to find balancing weights. With cross-fitting, our methods were shown to be asymptotically normal as the sample size grows for various regression learners, including neural networks and Lasso. For Lasso regression, we demonstrated that cross-fitting is unnecessary if the regularization parameters are chosen very carefully. However, this analysis also illustrates why cross-fitting is valuable in practice, as it is challenging to ensure the regularization parameters are chosen appropriately. Finally, we showcased the effectiveness of our proposed approach through computational experiments and a re-evaluation of the impact of minimum wage increases on youth employment.

\section*{Acknowledgement}

Research was sponsored by the Department of the Air Force Artificial Intelligence Accelerator and was accomplished under Cooperative Agreement Number FA8750-19-2-1000. The views and conclusions contained in this document are those of the authors and should not be interpreted as representing the official policies, either expressed or implied, of the Department of the Air Force or the U.S. Government. The U.S. Government is authorized to reproduce and distribute reprints for Government purposes notwithstanding any copyright notation herein.  This research was supported by NSF grants 1757140 and 224227.



\bibliographystyle{hapalike_mod}
\DeclareRobustCommand{\VAN}[2]{#2}  
\spacingset{1}{\bibliography{bib.bib}}
\spacingset{1.5}
\DeclareRobustCommand{\VAN}[2]{#1}  


\appendix
\section{Proof of Theorem~\ref{theorem:cross}}\label{sec:proof_cross}

\subsection{Asymptotic linearity}

   

To lighten notation, let $E_N(V)=N^{-1}\sum_{i=1}^NV_i$ and let $E_T(V)=T^{-1}\sum_{t=1}^TV_t$.

\begin{proof}[of Lemma~\ref{lemma:cross}]
We want to show
 $$\hat{\theta}=E_N \{m(Z,\gamma_0)\}+E_T[\alpha_0(X)\{Y-\gamma_0(X)\}]+o_p(N^{-1/2}).$$

By the continuous mapping theorem, it suffices to prove the result for within one fold, i.e. for
$$
\hat{\theta}_{\ell}=E_N \{m(Z,\hat{\gamma}_{\ell})\}+E_{T_{\ell}}[\hat{\alpha}_{\ell}(X)\{Y-\hat{\gamma}_{\ell}(X)\}].
$$

    To lighten notation, we hereafter we suppress the subscript $\ell$. We maintain the independence between $\hat{\alpha}$ and the training sample; and the independences among $\hat{\gamma}$, the training sample, and the target sample.

We proceed in steps.

\begin{enumerate}
    \item By algebraic manipulation, 
    $$
\hat{\theta}=E_N \{m(Z,\gamma_0)\}+E_T[\alpha_0(X)\{Y-\gamma_0(X)\}]+R_1+R_2+R_3
$$
where 
\begin{align*}
    R_1&=E_N\{m(Z,\hat{\gamma}-\gamma_0)\}+E_T[\alpha_0(X)\{\gamma_0(X)-\hat{\gamma}(X)\}] \\
    R_2&=E_T[\{\hat{\alpha}(X)-\alpha_0(X)\}\{Y-\gamma_0(X)\}] \\
    R_3&=E_T[\{\hat{\alpha}(X)-\alpha_0(X)\}\{\gamma_0(X)-\hat{\gamma}(X)\}].
\end{align*}
    \item To study $R_1$, define $\hat{\Delta}=\int m(z,\hat{\gamma}-\gamma_0)f(z)\mathrm{d}z$.
By Assumption~\ref{assumption:bounded},
    $
    \hat{\Delta}=\int \alpha_0(x)\{\hat{\gamma}(x)-\gamma_0(x)\}f(x)\mathrm{d}x.
    $
    We decompose $R_1=R_1-\hat{\Delta}+\hat{\Delta}=R_1'+R_1''$ where
    \begin{align*}
        R_1'&=E_N\{m(Z,\hat{\gamma}-\gamma_0)\}-\int m(z,\hat{\gamma}-\gamma_0)f(z)\mathrm{d}z \\
        R_1''&=E_T[\alpha_0(X)\{\gamma_0(X)-\hat{\gamma}(X)\}]-\int \alpha_0(x)\{\gamma_0(x)-\hat{\gamma}(x)\}f(x)\mathrm{d}x.
    \end{align*}
    Both $R_1'$ and $R_1''$ are mean zero by independence of $\hat{\gamma}$ from the target and training samples.
    Moreover, their second moments vanish quickly. 
    
    By the definition of $R_1'$, independent and identically distributed target observations in Assumption~\ref{assumption:bounded}, bounding the variance by the second moment, mean square continuity in Assumption~\ref{assumption:bounded}, and the regression rate in Assumption~\ref{assumption:learners},
    \begin{align*}
        E\{(R_1')^2 \mid \hat{\gamma}\}
        &=E([E_N\{m(Z,\hat{\gamma}-\gamma_0)\}-\int m(z,\hat{\gamma}-\gamma_0)f(z)\mathrm{d}z]^2 \mid \hat{\gamma}) \\
        &=\frac{1}{N}E(E_N[\{m(Z,\hat{\gamma}-\gamma_0)-\int m(z,\hat{\gamma}-\gamma_0)f(z)\mathrm{d}z\}^2]\mid \hat{\gamma}) \\
        &=\frac{1}{N}E[\{m(Z,\hat{\gamma}-\gamma_0)-\int m(z,\hat{\gamma}-\gamma_0)f(z)\mathrm{d}z)\}^2\mid \hat{\gamma}] \\
        &\leq \frac{1}{N}E[\{m(Z,\hat{\gamma}-\gamma_0)^2\mid \hat{\gamma}] 
        \leq \frac{C}{N} E[\{\hat{\gamma}(X)-\gamma_0(X)\}^2\mid \hat{\gamma}]
    =o_p(N^{-1}).
    \end{align*}

    By the definition of $R_1''$, independent and identically distributed training observations in Assumption~\ref{assumption:bounded}, bounding the variance by the second moment, $\|\alpha_0\|_{\infty}\leq \bar{\alpha}$, and the regression rate in Assumption~\ref{assumption:learners},
     \begin{align*}
        E\{(R_1'')^2 \mid \hat{\gamma}\}
        &=E\{ (E_T[\alpha_0(X)\{\gamma_0(X)-\hat{\gamma}(X)\}]-\int \alpha_0(x)\{\gamma_0(x)-\hat{\gamma}(x)\}f(x)\mathrm{d}x)^2 \mid \hat{\gamma}\} \\
        &=\frac{1}{T}
        E\{E_T(
        [\alpha_0(X)\{\gamma_0(X)-\hat{\gamma}(X)\}-\int \alpha_0(x)\{\gamma_0(x)-\hat{\gamma}(x)\}f(x)\mathrm{d}x]^2)
        \mid \hat{\gamma}\} \\
        &=\frac{1}{T}E([\alpha_0(X)\{\gamma_0(X)-\hat{\gamma}(X)\}-\int \alpha_0(x)\{\gamma_0(x)-\hat{\gamma}(x)\}f(x)\mathrm{d}x]^2\mid \hat{\gamma}) \\
        &\leq \frac{1}{T}E([\alpha_0(X)\{\gamma_0(X)-\hat{\gamma}(X)\}]^2\mid \hat{\gamma}) 
        \leq \frac{\bar{\alpha}^2}{T} E[\{\hat{\gamma}(X)-\gamma_0(X)\}^2\mid \hat{\gamma}]
    =o_p(T^{-1}).
    \end{align*}
    Therefore by the conditional Markov inequality and relative sample sizes in Assumption~\ref{assumption:bounded}, $R_1'=o_p(N^{-1/2})$ and $R_1''=o_p(N^{-1/2})$.
    
    \item To study $R_2$, observe that it is mean zero by independence of $\hat{\alpha}$ from the training samples. Moreover, its second moment vanishes quickly.
    By the definition of $R_2$, independent and identically distributed training observations in Assumption~\ref{assumption:bounded}, $\text{var}(Y\mid X)\leq\bar{\sigma}^2$, and the Riesz representer rate in Assumption~\ref{assumption:learners},
    \begin{align*}
        E(R_2^2 \mid \hat{\alpha} )
        &=E\{(E_T[\{\hat{\alpha}(X)-\alpha_0(X)\}\{Y-\gamma_0(X)\}])^2\mid\hat{\alpha}\} \\
        &=\frac{1}{T} E(E_T[\{\hat{\alpha}(X)-\alpha_0(X)\}^2\{Y-\gamma_0(X)\}^2]\mid\hat{\alpha}) \\
        &=\frac{1}{T} E[\{\hat{\alpha}(X)-\alpha_0(X)\}^2\{Y-\gamma_0(X)\}^2\mid\hat{\alpha}] \\
        &\leq  \frac{\bar{\sigma}^2}{T}E[\{\hat{\alpha}(X)-\alpha_0(X)\}^2\mid\hat{\alpha}]=o_p(T^{-1}).
    \end{align*}
    Therefore by the conditional Markov inequality and relative sample sizes in Assumption~\ref{assumption:bounded}, $R_2=o_p(N^{-1/2})$.
    
    \item To study $R_3$, observe that its first moment vanishes quickly. By the definition of $R_3$, the triangle inequality, independent and identically distributed training observations in Assumption~\ref{assumption:bounded}, the Cauchy-Schwarz inequality, and the joint rate condition in Assumption~\ref{assumption:learners},
    \begin{align*}
         E(|R_3|\mid \hat{\gamma},\hat{\alpha})
    &= E(|E_T[\{\hat{\alpha}(X)-\alpha_0(X)\}\{\gamma_0(X)-\hat{\gamma}(X)\}]|\mid \hat{\gamma},\hat{\alpha}) \\
    &\leq E(E_T[|\{\hat{\alpha}(X)-\alpha_0(X)\}\{\gamma_0(X)-\hat{\gamma}(X)\}|]\mid \hat{\gamma},\hat{\alpha}) \\ 
    &=E[|\{\hat{\alpha}(X)-\alpha_0(X)\}\{\gamma_0(X)-\hat{\gamma}(X)\}|\mid \hat{\gamma},\hat{\alpha}] \\
    &\leq (E[\{\hat{\alpha}(X)-\alpha_0(X)\}^2\mid \hat{\alpha}])^{1/2} (E[\{\gamma_0(X)-\hat{\gamma}(X)\}^2\mid \hat{\gamma}])^{1/2}
    =o_p(N^{-1/2}).
    \end{align*}
    Therefore by the conditional Markov inequality, $R_3=o_p(N^{-1/2})$.
\end{enumerate}
\end{proof}

\subsection{Trimming}

To lighten notation, we abbreviate $\tau(a)=\tau_N(a)$ and $\bar{\tau}=\bar{\tau}_N$.

\begin{lemma}[Trimming facts]\label{lemma:trim}
    The following facts hold under the last hypotheses of Theorem 1 and 2, where $1_{(\cdot)}$ is one if $(\cdot)$ is true and zero otherwise:
    $$
    |\tau(b)-\tau(a)|\leq |b-a|,\quad |\tau(a)-a|\leq 1_{|a|\geq\bar{\tau}}|a|,\quad E\{1_{|\alpha_0(X)|\geq\bar{\tau}}\alpha_0(X)^2\}=o(1).
    $$
\end{lemma}

\begin{proof}
    We prove each fact.
    \begin{enumerate}
        \item Without loss of generality, $b\geq a$. There are six cases. In three of the cases, $b\geq\bar{\tau}$. In two of the cases, $b\in (-\bar{\tau},\bar{\tau})$. In the final case, $b\leq\-\bar{\tau}$. It is straightforward to enumerate these cases on the real line and verify the desired inequality.
        \item There are three cases. Enumerating them on the real line confirms that
        $$
        |\tau(a)-a|\leq 1_{|a|\geq\bar{\tau}}(|a|-\bar{\tau})
        \leq 1_{|a|\geq\bar{\tau}}|a|.
        $$
        \item By Assumption~\ref{assumption:bounded}, $E\{\alpha_0^2(X)\}<\infty$. Then we appeal to the dominated convergence theorem.
    \end{enumerate}
\end{proof}

\begin{lemma}[Trimming consistency]\label{lemma:trim_converge}
    Suppose Assumptions~\ref{assumption:bounded} and~\ref{assumption:learners} hold. Then 
    $$
    E_{T_{\ell}}([\tau\{\hat{\alpha}_{\ell}(X)\}-\alpha_0(X)]^2)=o_p(1),\quad E_{T_{\ell}} [|\tau\{\hat{\alpha}_{\ell}(X)\}^2-\alpha_0(X)^2|]=o_p(1).
    $$
\end{lemma}

\begin{proof}
    To lighten notation, hereafter we suppress the subscript $\ell$. As in Lemma~\ref{lemma:cross}, we maintain the independence between $\hat{\alpha}$ and the training sample. We also write $\tilde{\alpha}(X)=\tau\{\hat{\alpha}(X)\}$. We prove each result.

    \begin{enumerate}
        \item Suppressing arguments, notice that
$$
\tilde{\alpha}-\alpha_0=\tau(\hat{\alpha})-\alpha_0=\{\tau(\hat{\alpha})-\tau(\alpha_0)\}-\{\tau(\alpha_0)-\alpha_0\}.
$$
Therefore by the parallelogram law and Lemma~\ref{lemma:trim},
    \begin{align*}
       (\tilde{\alpha}-\alpha_0)^2
        &\leq 2\{\tau(\hat{\alpha})-\tau(\alpha_0)\}^2+2\{\tau(\alpha_0)-\alpha_0\}^2 \\
        &\leq 2(\hat{\alpha}-\alpha_0)^2+2 1_{|\alpha_0|\geq\bar{\tau} }|\alpha_0|^2.
    \end{align*}
    Therefore it suffices to show
    \begin{equation}\label{eq:alpha}
         E_T [\{\hat{\alpha}(X)-\alpha_0(X)\}^2]=o_p(1)
    \end{equation}
    and
$
E_T\{1_{|\alpha_0(X)|\geq\bar{\tau} }\alpha_0(X)^2\}=o_p(1).
$
    The former holds by the conditional Markov inequality. In particular, its first moment vanishes by Assumptions~\ref{assumption:bounded} and~\ref{assumption:learners}:
   \begin{align*}
        E (| E_T [\{\hat{\alpha}(X)-\alpha_0(X)\}^2] | \mid \hat{\alpha})
        &=
         E  (E_T [\{\hat{\alpha}(X)-\alpha_0(X)\}^2]  \mid \hat{\alpha}) \\
         &=  E  [\{\hat{\alpha}(X)-\alpha_0(X)\}^2 \mid \hat{\alpha}]
         =o_p(1).
   \end{align*}
   The latter holds by the Markov inequality. In particular, its first moment vanishes by Assumption~\ref{assumption:bounded} and Lemma~\ref{lemma:trim}:
   $$
   E [|E_T\{1_{|\alpha_0(X)|\geq\bar{\tau} }\alpha_0(X)^2\}|]
   = E [E_T\{1_{|\alpha_0(X)|\geq\bar{\tau} }\alpha_0(X)^2\}]
   =  E \{1_{|\alpha_0(X)|\geq\bar{\tau} }\alpha_0(X)^2\}
   =o(1).
   $$
        \item Suppressing arguments, write
        \begin{align*}
            E_T(|\tilde{\alpha}^2-\alpha_0^2|)
            &=E_T\{|(\tilde{\alpha}-\alpha_0)(\tilde{\alpha}+\alpha_0)|\} 
            \leq [E_T\{(\tilde{\alpha}-\alpha_0)^2\}]^{1/2}[E_T\{(\tilde{\alpha}+\alpha_0)^2\}]^{1/2}.
        \end{align*}
        By the previous result, the former factor is $o_p(1)$.
        Focusing on the latter factor, by the parallelogram law,
        \begin{align*}
            E_T\{(\tilde{\alpha}+\alpha_0)^2\}
            &=E_T\{(\tilde{\alpha}-\alpha_0+2\alpha_0)^2\} 
            \leq 2E_T\{(\tilde{\alpha}-\alpha_0)^2\}+2E_T \{(2\alpha_0)^2\}.
        \end{align*}
        By the previous result, the first term is $o_p(1)$. By the weak law of large numbers, the second term is $O_p(1)$ when $E(\alpha_0^2)<\infty$, which holds by Assumption~\ref{assumption:bounded}.
        In summary, as desired,
        $$
        E_T(|\tilde{\alpha}^2-\alpha_0^2|)=o_p(1)\{o_p(1)+O_p(1)\}=o_p(1).
        $$
    \end{enumerate}

\end{proof}

\subsection{Main result}

\begin{proof}[of Theorem~\ref{theorem:cross}]
    We proceed in steps.
    \begin{enumerate}
        \item For asymptotic normality, we appeal to Lemma~\ref{lemma:cross} and the central limit theorem. In particular, the defining formula gives $\theta_0=E\{m(Z,\gamma_0)\}$ and the law of iterated expectations gives $E[\alpha_0\{Y-\gamma_0\}]=0$ so
        \begin{align*}
            N^{1/2} \{E_N\{m(Z,\gamma_0)\}-\theta_0\}
            &\rightsquigarrow \mathcal{N}[0,\text{var}\{m(Z,\gamma_0\}] \\ 
            T^{1/2} E_T[\alpha_0(X)\{Y-\gamma_0(X)\}]
            &\rightsquigarrow \mathcal{N}(0,E[\alpha_0(X)^2\{Y-\gamma_0(X)\}^2]).
        \end{align*}
The two empirical processes are independent by Assumption~\ref{assumption:bounded}. Finally, $(N/T)^{1/2}$ times the latter process has the asymptotic distribution $\mathcal{N}(0,\xi E[\alpha_0(X)^2\{Y-\gamma_0(X)\}^2])$. In summary,
$$
\sigma^2=\text{var}\{m(Z,\gamma_0)\}+\xi E\{\alpha_0(X)^2\text{var}(Y\mid X)\}.
$$
        \item For variance estimation, it suffices to prove the result within one fold, i.e. for 
        $$
        \hat{\sigma}_{\ell}^2=E_N \{m(Z,\hat{\gamma}_{\ell})^2\}-[E_N \{m(Z,\hat{\gamma}_{\ell})\}]^2+\frac{N}{T_{\ell}}E_{T_{\ell}} [\tau\{\hat{\alpha}_{\ell}(X)\}^2\{Y-\hat{\gamma}_{\ell}(X)\}^2].
        $$
        To lighten notation, hereafter we suppress the subscript $\ell$. As in Lemma~\ref{lemma:cross}, we maintain the independence between $\hat{\alpha}$ and the training sample; and the independences among $\hat{\gamma}$, the training sample, and the target sample.
        \item To begin, we show 
        $$
E_N \{m(Z,\hat{\gamma})^2\}-[E_N \{m(Z,\hat{\gamma})\}]^2=\text{var}\{m(Z,\gamma_0)\}+o_p(1).
        $$
       Define the notation 
       \begin{align*}
           \hat{\psi}(Z)&=m(Z,\hat{\gamma})-E_N\{m(Z,\hat{\gamma})\},\quad 
           \psi_0(Z)=m(Z,\gamma_0)-E\{m(Z,\gamma_0)\}.
       \end{align*}
       Further define
       \begin{align*}
           \hat{\sigma}_m^2&=E_N \{m(Z,\hat{\gamma})^2\}-[E_N \{m(Z,\hat{\gamma})\}]^2=E_N\{\hat{\psi}(Z)^2\} \\
           \sigma_m^2&=\text{var}\{m(Z,\gamma_0)\}=E\{\psi_0(Z)^2\}.
       \end{align*}
By \citet[Theorem 3]{chernozhukov2023simple},
$$
|\hat{\sigma}_m^2-\sigma_m^2|\leq \Delta_m'+2(\Delta_m')^{1/2}\{(\Delta_m'')^{1/2}+\sigma_m\}+\Delta_m''
$$
where $(\Delta_m',\Delta_m'')$ are given below. Therefore by the continuous mapping theorem, it suffices to show
\begin{align*}
    \Delta_m'&=E_N[\{\hat{\psi}(Z)-\psi_0(Z)\}^2]=o_p(1),\quad
    \Delta_m''=|E_N\{\psi_0(Z)^2\}-\sigma_m^2|=o_p(1).
\end{align*}
The latter is immediate by the weak law of large numbers  if $E\{\psi_0(Z)^2\}<\infty$.
For the former, observe that
\begin{align*}
    \hat{\psi}(Z)-\psi_0(Z)
    &=m(Z,\hat{\gamma})-E_N\{m(Z,\hat{\gamma})\} -m(Z,\gamma_0)+E\{m(Z,\gamma_0)\}\\
    &=m(Z,\hat{\gamma}-\gamma_0)+E\{m(Z,\gamma_0)\}-E_N\{m(Z,\gamma_0)\}+E_N\{m(Z,\gamma_0-\hat{\gamma})\}.
\end{align*}
Therefore by the parallelogram law,
\begin{align*}
    \Delta_m'\leq 3E_N\{m(Z,\hat{\gamma}-\gamma_0)^2\}+3[E\{m(Z,\gamma_0)\}-E_N\{m(Z,\gamma_0)\}]^2+3 E_N ([E_N\{m(Z,\gamma_0-\hat{\gamma})\}]^2).
\end{align*}
The middle expression is $o_p(1)$ by the weak law of large numbers. By Jensen's inequality, the third expression is upper bounded by the first expression. The first expression is $o_p(1)$ by the conditional Markov inequality since its first moment is
$$
E[E_N\{m(Z,\hat{\gamma}-\gamma_0)^2\}\mid\hat{\gamma }]=
E\{m(Z,\hat{\gamma}-\gamma_0)^2\mid\hat{\gamma}\} 
\leq C E[ \{\hat{\gamma}(X)-\gamma_0(X)\}^2 \mid\hat{\gamma} ]=o_p(1),
$$
  where we appeal to Assumptions~\ref{assumption:bounded} and~\ref{assumption:learners}.   
        \item Next, we show
        $$
        E_T [\tau\{\hat{\alpha}(X)\}^2\{Y-\hat{\gamma}(X)\}^2]=E\{\alpha_0(X)^2\text{var}(Y\mid X)\}+o_p(1).
        $$
Then we are done, since $N/T$ converges to $\xi$ by Assumption~\ref{assumption:bounded}.
        
Consider the decomposition
$$
E_T [\tau\{\hat{\alpha}(X)\}^2\{Y-\hat{\gamma}(X)\}^2]-E\{\alpha_0(X)^2\text{var}(Y\mid X)\}
=S_1+S_2+S_3
$$
where
\begin{align*}
    S_1&=E_T (\tau\{\hat{\alpha}(X)\}^2[\{Y-\hat{\gamma}(X)\}^2-\{Y-\gamma_0(X)\}^2]) \\
    S_2&=E_T ([\tau\{\hat{\alpha}(X)\}^2-\alpha_0(X)^2]\{Y-\gamma_0(X)\}^2) \\
    S_3&=E_T[\alpha_0(X)^2\{Y-\gamma_0(X)\}^2]- E\{\alpha_0(X)^2\text{var}(Y\mid X)\}.
\end{align*}
When analyzing $(S_1,S_2,S_3)$, we lighten notation as
$$
\epsilon=Y-\gamma_0(X),\quad \delta=\hat{\gamma}(X)-\gamma_0(X),\quad \tilde{\alpha}(X)=\tau\{\hat{\alpha}(X)\}.
$$
\begin{enumerate}
    \item To study $S_1$, write each summand as
    \begin{align*}
        \tilde{\alpha}(X)^2\{(\epsilon-\delta)^2-\epsilon^2\}
        =
        \tilde{\alpha}(X)^2\{-2\epsilon\delta+\delta^2\}
        =-2\epsilon\delta \tilde{\alpha}(X)^2
        + \delta^2\tilde{\alpha}(X)^2.
    \end{align*}
    Therefore it suffices to show
    $$
    E_T\{\epsilon\delta \tilde{\alpha}(X)^2\}=o_p(1),\quad E_T\{\delta^2\tilde{\alpha}(X)^2\}=o_p(1).
    $$
\begin{enumerate}
    \item Consider the former expression. By Cauchy-Schwarz inequality,
    \begin{align*}
        E_T\{\epsilon\delta \tilde{\alpha}(X)^2\} 
        &\leq \{E_T(\delta^2)\}^{1/2}
        [E_T\{\epsilon^2\tilde{\alpha}(X)^4\}]^{1/2}.
    \end{align*}
Moreover, 
\begin{equation}\label{eq:gamma}
    E_T(\delta^2)=E_T[\{\hat{\gamma}(X)-\gamma_0(X)\}^2]=o_p(\bar{\tau}^{-2})
\end{equation}
by the conditional Markov inequality, since its first moment is, using Assumption~\ref{assumption:bounded} and the gradual trimming condition, $$
E\{|E_T(\delta^2)| \mid\hat{\gamma}\}
=E\{E_T(\delta^2) \mid\hat{\gamma}\}
=E(\delta^2 \mid\hat{\gamma})
=o_p(\bar{\tau}^{-2}).
$$
Finally we show $E_T\{\epsilon^2\tilde{\alpha}(X)^4\}=O_p(\bar{\tau}^2)$ using the conditional Markov inequality. Its first moment is, by Assumption~\ref{assumption:bounded}, $\text{var}(Y\mid X)\leq\bar{\sigma}^2$, and trimming,
\begin{align*}
    E [|E_T\{\epsilon^2\tilde{\alpha}(X)^4\}| \mid \hat{\alpha} ]
    &= E[ E_T\{\epsilon^2\tilde{\alpha}(X)^4\} \mid \hat{\alpha}]  \\
    &= E \{\epsilon^2\tilde{\alpha}(X)^4 \mid \hat{\alpha}\}  \\
    &\leq \bar{\sigma}^2 E \{\tilde{\alpha}(X)^4 \mid \hat{\alpha}\} \\
    &\leq \bar{\sigma}^2 \bar{\tau}^2 E \{\tilde{\alpha}(X)^2 \mid \hat{\alpha}\}.
\end{align*}
Moreover, by the parallelogram law, Lemma~\ref{lemma:trim_converge}, and Assumption~\ref{assumption:bounded}
\begin{align*}
    E \{\tilde{\alpha}(X)^2 \mid \hat{\alpha}\}
&=E [\{\tilde{\alpha}(X)-\alpha_0(X)+\alpha_0(X)\}^2 \mid \hat{\alpha}] \\
&\leq 2E [\{\tilde{\alpha}(X)-\alpha_0(X)\}^2 \mid \hat{\alpha}]
+2E \{\alpha_0(X)^2 \mid \hat{\alpha}\} \\
&=o_p(1)+O(1).
\end{align*}
In summary, as desired,
$$
E_T\{\epsilon\delta \tilde{\alpha}(X)^2\}=o_p(\bar{\tau}^{-1})O_p(\bar{\tau})=o_p(1).
$$
    \item Consider the latter expression. By trimming and $E_T(\delta^2)=o_p(\bar{\tau}^{-2})$ as previously argued,
    $$
    E_T\{\delta^2\tilde{\alpha}(X)^2\}\leq \bar{\tau}^2 E_T(\delta^2)=\bar{\tau}^2o_p(\bar{\tau}^{-2})=o_p(1).
    $$
\end{enumerate}
\item To study $S_2$, write each summand as 
$
\{\tilde{\alpha}(X)^2-\alpha_0(X)^2\}\epsilon^2.
$ 
Once again we use the conditional Markov inequality. Its first moment is, by the definition of $S_2$, Assumption~\ref{assumption:bounded}, $\text{var}(Y\mid X)\leq \bar{\sigma}^2$, and Lemma~\ref{lemma:trim_converge},
\begin{align*}
    E\{|S_2|\mid (X_t),(Z_i)\}
    &=E(|E_T[\{\tilde{\alpha}(X)^2-\alpha_0(X)^2\}\epsilon^2]|\mid (X_t),(Z_i)) \\
    &\leq E[E_T\{|\tilde{\alpha}(X)^2-\alpha_0(X)^2|\epsilon^2\}\mid (X_t),(Z_i)] \\
    &=E_T\{|\tilde{\alpha}(X)^2-\alpha_0(X)^2|\} E\{\epsilon^2\mid (X_t),(Z_i)\} \\
    &\leq \bar{\sigma}^2 E_T\{|\tilde{\alpha}(X)^2-\alpha_0(X)^2|\}=o_p(1).
\end{align*}

\item To study $S_3$, we appeal to the law of large numbers and the fact that $E[\alpha_0(X)^2\{Y-\gamma_0(X)\}^2]\leq \bar{\sigma}^2 E\{\alpha_0(X)^2\}$, and the final expression is finite by Assumption~\ref{assumption:bounded}. 
\end{enumerate}
        
    \end{enumerate}
\end{proof}

\section{Proof of Theorem~\ref{theorem:lasso}}

We abbreviate notation as in Section~\ref{sec:proof_cross}.

\begin{proof}[of Theorem~\ref{theorem:lasso}]
    We appeal to the arguments proving \citet[Lemmas B11 and B13]{bradic2022minimax}. The correspondence between our assumptions and theirs is as follows. 
    
    Under the regularity conditions $|\alpha_0(X)|\leq\bar{\alpha}$, $\text{var}(Y\mid X)\leq\bar{\sigma}^2$, and $E(Y^2)<\infty$,
    Assumption~\ref{assumption:correct} verifies \citet[Assumption 8]{bradic2022minimax},
    Assumption~\ref{assumption:regular} matches \citet[Assumption 4]{bradic2022minimax} with $\tilde{W}=(Z,X)$,
    Assumption~\ref{assumption:dictionary} matches \citet[Assumption 6]{bradic2022minimax},
    Assumption~\ref{assumption:regularization} matches \citet[Assumption 5]{bradic2022minimax},
    Assumption~\ref{assumption:sparse} matches \citet[Assumptions 3 and 9(i)]{bradic2022minimax},
    Assumptions~\ref{assumption:bounded} and~\ref{assumption:approx} verify \citet[Assumptions 9(ii) and 9(iii)a]{bradic2022minimax}, and
    Assumption~\ref{assumption:eigenvalue} matches \citet[Assumption 7]{bradic2022minimax}.

    In summary, we satisfy the hypotheses of \citet[Corollary 9]{bradic2022minimax}. This result is not a consequence of \citet[Corollary 9]{bradic2022minimax} due to the data fusion of training and target data, yet we are able to adapt their intermediate results.

    Recall that $(\hat{\beta},\hat{\rho})$ are the coefficients of $(\hat{\gamma},\hat{\alpha})$.
    Let $(\beta_N,\rho_N)$ be the coefficients of $(\gamma_N,\alpha_N)$, and define $\theta_N=\beta_N^{\top}Q \rho_N$. Let $\hat{\mu}=E_T\{Yb(X)\}$, $\hat{U}=\hat{\mu}-\hat{Q}\beta_N$, and $\hat{R}=\hat{M}-\hat{Q}\rho_N$. 

    We proceed in steps.

    \begin{enumerate}
        \item  To begin, we study the sampling error. By \citet[eq. B.4]{bradic2022minimax},
    $
    \hat{\theta}-\theta_N=\hat{\Psi}_N+T_1+T_2+T_3
    $
    where
    \begin{align*}
        \hat{\Psi}_N&=E_N\{m(Z,\gamma_N)\}-\theta_N+E_T[\alpha_N(X)\{Y-\gamma_N(X)\}] \\
        T_1&=(\beta_N-\hat{\beta})^{\top}\hat{Q}(\hat{\rho}-\rho_N),\quad 
        T_2=\hat{R}^{\top}(\hat{\beta}-\beta_N),\quad 
        T_3=\hat{U}^{\top}(\hat{\rho}-\rho_N).
    \end{align*} 
    
By Assumption~\ref{assumption:bounded}, $N$ and $T$ are asymptotically proportional. Therefore by the proof of \citet[Lemma B.11]{bradic2022minimax}, $T_1=o_p(N^{-1/2})$ and $T_2=o_p(N^{-1/2})$.

    We further decompose $T_3=T_3'+T_3''$ where $\epsilon=Y-\gamma_0(X)$, $D=\gamma_0(X)-\gamma_N(X)$, and hence
    $$
    T_3'=E_T\{\epsilon b(X)^{\top}(\hat{\rho}-\rho_N)\},\quad T_3''=E_T\{Db(X)^{\top}(\hat{\rho}-\rho_N)\}.
    $$
    By independence of training and target samples in Assumption~\ref{assumption:bounded}, $$
    E\{T_3'|(X_t),(Z_i)\}=0,\quad \text{var}\{T_3'|(X_t),(Z_i)\}\leq \frac{\bar{\sigma}^2}{T}(\hat{\rho}-\rho_N)^{\top}\hat{Q}(\hat{\rho}-\rho_N)=o_p(N^{-1})$$
    where the final statement appeals to \citet[Lemma B.9]{bradic2022minimax}. Therefore by the conditional Markov inequality,
    $T_3'=o_p(N^{-1/2})$. As in the proof of \citet[Lemma B.11]{bradic2022minimax}, $T_3''=o_p(N^{-1/2})$. In summary, $T_3=o_p(N^{-1/2})$ by the triangle inequality.

Collecting results, $\hat{\theta}-\theta_N=\hat{\Psi}_N+o_p(N^{-1/2})$.

    \item Next we prove asymptotic linearity.
    Define
    $$
    \hat{\Psi}=E_N\{m(Z,\gamma_0)\}-\theta_0+E_T[\alpha_0(X)\{Y-\gamma_0(X)\}].
    $$
    Similar to \citet[Lemma B.13]{bradic2022minimax}, $\hat{\Psi}_N+\theta_N-\hat{\Psi}-\theta_0=o_p(N^{-1/2})$. Therefore by the result in the previous step and the result in this step,
    $$
    N^{1/2}(\hat{\theta}-\theta_0)
    =N^{1/2}(\hat{\theta}-\theta_N+\theta_N-\theta_0)
    =N^{1/2}(\hat{\Psi}_N+\theta_N-\theta_0)+o_p(1)
    =N^{1/2}\hat{\Psi}+o_p(1).
    $$
    Asymptotic normality follows from the central limit and Slutsky theorems.

    \item The argument for variance estimation mirrors the proof of Theorem~\ref{theorem:cross}. We prove that 
    $$
 E_T[\{\hat{\alpha}(X)-\alpha_0(X)\}^2]=o_p(1),\quad E_T[\{\hat{\gamma}(X)-\gamma_0(X)\}^2]=o_p(\bar{\tau}_N^{-2}).
$$
Thereafter, the argument is identical, using these justifications instead of the previous justifications for equations~\eqref{eq:alpha} and~\eqref{eq:gamma}, respectively. To lighten notation, let $\varepsilon_N=\{\ln(J)/N\}^{1/2}$.

    By the parallelogram law, 
    \begin{align*}
        E_T[\{\hat{\alpha}(X)-\alpha_0(X)\}^2]
        &=
        E_T[\{\hat{\alpha}(X)-\alpha_N(X)+\alpha_N(X)\alpha_0(X)\}^2] \\
        &\leq 2(\hat{\rho}-\rho_N)^{\top}\hat{Q}(\hat{\rho}-\rho_N)+2E_T[\{\alpha_N(X)-\alpha_0(X)\}^2] \\
        &=o_p(1).
    \end{align*}
    The former term is $o_p(1)$ by \citet[Lemma B.9]{bradic2022minimax}. The latter is $o_p(1)$ by the law of large numbers and Assumption~\ref{assumption:approx}.

Similarly, by the parallelogram law, H\"older inequality, and Assumption~\ref{assumption:dictionary}
\begin{align*}
    E_T[\{\hat{\gamma}(X)-\gamma_0(X)\}^2]
    &=
     E_T[\{\hat{\gamma}(X)-\gamma_N(X)+\gamma_N(X)-\gamma_0(X)\}^2]\\
     &\leq 2(\hat{\beta}-\beta_N)^{\top}\hat{Q}(\hat{\beta}-\beta_N)+2E_T[\{\gamma_N(X)-\gamma_0(X)\}^2] \\
    &\leq 2\bar{b}^2 \|\hat{\beta}-\beta_N\|_1^2+2E_T[\{\gamma_N(X)-\gamma_0(X)\}^2].
\end{align*}
Using \citet[Lemmas B.4 and B.7]{bradic2022minimax} and that Assumption~\ref{assumption:approx}(ii) implies $r\varepsilon_N^{-1}=o(N^c)$ for every $c>0$,
    $$
    \|\hat{\beta}-\beta_N\|_1=O_p \left\{ r \varepsilon_N^{-1} \varepsilon_N^{(2\zeta-1)/(2\zeta+1)}\right\}=O_p \left\{ N^c \varepsilon_N^{(2\zeta-1)/(2\zeta+1)}\right\}.
    $$
By the law of large numbers, 
$$
E_T[\{\gamma_N(X)-\gamma_0(X)\}^2]=E[\{\gamma_N(X)-\gamma_0(X)\}^2]+o_p(1).
$$
  Finally recall the hypothesis that for some $c>0$,
$$
 N^c (\varepsilon_N^2)^{(2\zeta-1)/(2\zeta+1)} +E[\{\gamma_N(X)-\gamma_0(X)\}^2]=o(\bar{\tau}_N^{-2}).
$$
    \end{enumerate}
\end{proof}

\section{Simulation details}

Replication code for our analysis can be found in a public \href{https://github.com/vsyrgkanis/autodml-shifts}{GitHub repository}.
We implement a neural network with four hidden layers of 32 nodes each and rectified linear unit (ReLU) activation.
 We employ \texttt{MATLAB}'s \texttt{trainNetwork} function with \texttt{featureInputLayer} and \texttt{fullyConnectedLayer} settings, a learning rate of $0.01$, a mini-batch size of $1,024$, and up to $500$ epochs of training, applying $L_2$ regularization of $0.0002$.  We use the adaptive moment estimation (Adam) optimizer.






\end{document}